\documentclass[aps,prd,showpacs,amssymb,nofootinbib,draft]{revtex4}
\usepackage{graphicx}
\newcommand{\be}{\begin{eqnarray*}}
\newcommand{\ee}{\end{eqnarray*}}
\begin{document}
\title{Lorentz Violations and Euclidean Signature Metrics}

\author{J. Fernando \surname{Barbero G.}}
\email[]{jfbarbero@imaff.cfmac.csic.es} \affiliation{Instituto de
Matem\'aticas y F\'{\i}sica Fundamental, Centro de F\'{\i}sica
Miguel A. Catal\'{a}n, C.S.I.C., Serrano 113bis, 28006 Madrid,
Spain}
\author{Eduardo J. \surname{S. Villase\~nor}}
\email[]{eduardo@imaff.cfmac.csic.es} \affiliation{Instituto de
Matem\'aticas y F\'{\i}sica Fundamental, Centro de F\'{\i}sica
Miguel A. Catal\'{a}n, C.S.I.C., Serrano 113bis, 28006 Madrid,
Spain}

\date{July 14, 2003}

\begin{abstract}
We show that the families of effective actions considered by
Jacobson \textit{et al.} to study Lorentz invariance violations
contain a class of models that represent pure General Relativity
with Euclidean signature. We also point out that some members of
this family of actions preserve Lorentz invariance in a
generalized sense.
\end{abstract}


\maketitle

In recent years there have been several proposals to study Lorentz
invariance violations in general relativity and their
observational consequences (see
\cite{Jacobson:2001yj},\cite{Jacobson:2003ty},\cite{Mattingly:2001yd}
and references therein). The main ingredient of these models is
the introduction of a preferred frame (referred to by the authors
as the \textit{aether}) described by a unit timelike vector field
$u^a$. In order to preserve general covariance $u^a$ is taken as a
dynamical field. The most general action considered in these
papers has the form
\begin{eqnarray}
{\cal L}_{g,u} &=& a_0 - a_1 R - a_2 R_{ab}u^au^b - b_1\,
F^{ab}F_{ab}  \nonumber\\
& & -b_2\, (\nabla_a u_b) (\nabla^a u^b) -b_3\,
\dot{u}^a\dot{u}_a+\lambda(g^{ab}u_au_b-1), \label{Lgu}
\end{eqnarray}
where $\dot{u}^a:= u^m\nabla_m u^a$, $\lambda$ is the Lagrange
multiplier that enforces the condition that $u^a$ is a unit
vector, and $F_{ab}$ is defined as $F_{ab}:= 2\nabla_{[a} u_{b]}$.
It is important to notice that the models described by (\ref{Lgu})
are not the usual tensor-vector theories due to this constraint.
This type of Lagrangians have been already considered in the
literature by Kosteleck\'y and Samuel \cite{Kostelecky:1989jw} for
gravitational models and by Kosteleck\'y and Mewes
\cite{Kostelecky:2002hh} in the context of electrodynamics. The
role of questions similar to the ones discussed here, in
particular coordinate invariance, in the construction of
dispersion relations with physical Lorentz violation is discussed
in \cite{Lehnert:2003ue}.

\bigskip

We want to point out here that some of these actions can be
interpreted as describing \textit{pure} general relativity with
Euclidean signature and others are, in fact, equivalent to
Lorentzian general relativity without any Lorentz violating
effects.

Following the ideas presented in \cite{Barbero:1996ud} let us
consider a metric
\begin{equation}
g^{E}_{ab}=-\frac{1}{2}\sqrt{|\alpha(\alpha+2\beta)|}\left
[g_{ab}-2\frac{\alpha+\beta}{\alpha+2\beta}u_{a}u_{b}\right]
\label{f1}
\end{equation}
where $\alpha$ and $\beta$ are two real parameters, $g_{ab}$ is a
Lorentzian metric [with $(+---)$ signature], and $u^a$ is a unit
timelike vector field $(g_{ab}u^a u^b=1)$. Here $u_a\equiv
g_{ab}u^a$. If we compute the determinant of $g^{E}_{ab}$ we
obtain
\begin{equation}
g^E\equiv\det g^{E}_{ab}=-\frac{1}{16}\alpha^3 (\alpha+2\beta)\det
g_{ab}.
\end{equation}
As we can see the fact that $u^a$ is a unit vector implies that
the determinants of $g^{E}_{ab}$ and $g_{ab}$ are proportional to
each other with a constant of proportionality that can be made
either positive or negative by choosing appropriate values of
$\alpha$ and $\beta$. Let us write now the Einstein-Hilbert action
for $g^{E}_{ab}$ as a function of $g_{ab}$ and $u^a$. To this end
we need the inverse metric $g^{E\,ab}$ and the Christoffel symbols
for $g^{E}_{ab}$ (here $g^{ab}$ satisfies
$g_{ab}g^{bc}=\delta_a^c$)
\begin{eqnarray}
 g^{E\,ab}&=&-\frac{2}{\sqrt{|\alpha(\alpha+2\beta)|}}\left[
g^{ab}-2\frac{\alpha+\beta}{\alpha}u^{a}u^{b}\right]\,,\label{f2}
\\
\Gamma^{Ea}_{\,\,\,\,\,bc}&=&\Gamma^{a}_{bc}-\frac{\alpha+\beta}{\alpha+2\beta}
\left[\nabla_b(u^a u_c)+\nabla_c(u^a u_b)-\nabla^a(u_b
u_c)\right]\nonumber\\
& &
+\frac{2(\alpha+\beta)^2}{\alpha(\alpha+2\beta)}\left[u^a\nabla_b
u_c+u^a\nabla_c u_b-u^a u^d \nabla_d(u_b u_c)\right]\label{f3}\,.
\end{eqnarray}
A tedious but straightforward computation gives now
\begin{eqnarray}
S_E=\int d^4x\,\sqrt{|g_E|}g^{E\,ab}R^E_{ab}\hspace{8cm}& &\nonumber\\
={\rm sgn}(\alpha)\int
d^4x\,\sqrt{|g|}\left[-\frac{\alpha}{2}R+(\alpha+\beta)u^a
u^bR_{ab}-\frac{(\alpha+\beta)^2}
{\alpha+2\beta}g^{ab}\omega_a\omega_b\right] \label{f4}
\end{eqnarray}
where $R^E_{ab}$ is the Ricci tensor\footnote{Throughout this
paper we are using the conventions of Wald \cite{Wald:1984rg} for
the definitions of geometric objects and, in particular, for the
Riemann tensor.} built with $g^{E}_{ab}$, $R_{ab}$ and $R$ with
$g_{ab}$, and $\omega_a$ is the twist of $u^a$ given by
\begin{equation}
\omega_a=\epsilon_{a a_1 a_2
a_3}u^{a_1}\nabla^{a_2}u^{a_3}.\label{f5}
\end{equation}
It is useful to notice that
\begin{equation}
\omega_a \omega^a=(\nabla_a u_b)(\nabla^a u^b)-(\nabla_a
u_b)(\nabla^b
u^a)-\dot{u}^a\dot{u}_a=\frac{1}{2}F_{ab}F^{ab}-\dot{u}^a\dot{u}_a.\label{f6}
\end{equation}
As in (\ref{Lgu}) the condition that $u^a$ is a unit vector can be
explicitly incorporated to the action by adding a suitable
Lagrange multiplier term to (\ref{f4}). Another way to do that
\cite{Barbero:1996ud} is to write $u^a=\eta^a/(g_{bc}\eta^b
\eta^c)^{1/2}$, with an unconstrained, time-like, vector field
$\eta^a$; in which case the action becomes invariant under the
gauge transformations consisting in local rescalings of the vector
field. We can readily see that (\ref{f4}) is a particular case of
the action (\ref{Lgu}) considered in \cite{Jacobson:2001yj} with
the parameter choices $a_0=0$, $a_1=|\alpha/2|$, $a_2=-{\rm
sgn}(\alpha)(\alpha+\beta)$, $b_1={\rm
sgn}(\alpha)\frac{(\alpha+\beta)^2}{2(\alpha+2\beta)}$, $b_2=0$,
and $b_3=-2 b_1$.

\bigskip

\noindent Several comments are now in order:

\begin{itemize}

\item[i)] Some of the parameter choices do not change the
signature of the metric. If both $g^E_{ab}$ and $g_{ab}$ have
Lorentzian signatures the action (\ref{f4}) is strictly equivalent
to the Einstein-Hilbert action for $g_{ab}$. It is important to
realize that (\ref{f4}) has a gauge symmetry that is related to
the fact that the variations in the vector field can always be
compensated inside $g^E_{ab}$ by a suitable variation of the
metric $g_{ab}$. This also means that the field equations coming
from variations in the vector field are always redundant. We see
then that there is a one to one correspondence between the
solutions to the field equations for the Einstein-Hilbert action
(Lorentzian or Euclidean) and gauge equivalence classes of
solutions to the field equations derived from (\ref{f4}). In our
opinion it would not be justified to talk about Lorentz violating
effects when $g^E_{ab}$ is Lorentzian.

\item[ii)] The fact that $u^a$ is dynamical or not is irrelevant
in our scheme. If $u^a$ is a fixed geometric structure, general
covariance is broken but, as long as matter couples to the
$g^E_{ab}$, the physical content of the model corresponds to
general relativity in the sense that there is a one to one
correspondence between the solutions of the two theories and their
symmetries.

\item[iii)] Due to the presence of two different metrics
$g^E_{ab}$ and $g_{ab}$ one can consider matter couplings to any
of them. If matter is coupled to $g^E_{ab}$ and the parameters of
the model are chosen in such a way that $g^E_{ab}$  is Lorentzian
we still have Lorentzian general relativity \textit{without
breaking} any Lorentz invariance in the sense discussed above. If,
on the other hand, we choose the parameters to get Euclidean
signature we would end up with a Euclidean general relativity with
matter. Finally, if matter is coupled to $g_{ab}$ we would have
the Lorentz violating effects described in
\cite{Jacobson:2001yj},\cite{Jacobson:2003ty},\cite{Mattingly:2001yd}.

\item[iv)] The field equations obtained by varying in $u^a$ are
redundant. This can be explicitly checked by varying our action
(\ref{f4}) with respect to $u^a$ and checking that the equations
thus obtained are satisfied as a consequence of the equations
derived by varying with respect to $g_{ab}$. This can also be seen
by noticing that a variation in $g^E_{ab}$ of the type generated
by changing $u^a$ can be obtained, also, by a suitable variation
of $g_{ab}$, as discussed above.

\item[v)] If $u^{a}$ is hypersurface orthogonal the twist is not
present and we get the formulation presented in
\cite{Barbero:1996ud} in the context of real Wick rotations.

\end{itemize}

\begin{acknowledgments} The authors wish to thank G. Mena and T.
Jacobson for valuable comments. This work was supported by the
Spanish MCYT under the research project BFM2002-04031-C02-02.
E.J.S.V. is supported by a Spanish Ministry of Education and
Culture fellowship co-financed by the European Social Fund.
\end{acknowledgments}

%\bibliography{signat}

\begin{thebibliography}{8}
\expandafter\ifx\csname
natexlab\endcsname\relax\def\natexlab#1{#1}\fi
\expandafter\ifx\csname bibnamefont\endcsname\relax
  \def\bibnamefont#1{#1}\fi
\expandafter\ifx\csname bibfnamefont\endcsname\relax
  \def\bibfnamefont#1{#1}\fi
\expandafter\ifx\csname citenamefont\endcsname\relax
  \def\citenamefont#1{#1}\fi
\expandafter\ifx\csname url\endcsname\relax
  \def\url#1{\texttt{#1}}\fi
\expandafter\ifx\csname
urlprefix\endcsname\relax\def\urlprefix{URL }\fi
\providecommand{\bibinfo}[2]{#2}
\providecommand{\eprint}[2][]{\url{#2}}

\bibitem[{\citenamefont{Jacobson and Mattingly}(2001)}]{Jacobson:2001yj}
\bibinfo{author}{\bibfnamefont{T.}~\bibnamefont{Jacobson}} \bibnamefont{and}
  \bibinfo{author}{\bibfnamefont{D.}~\bibnamefont{Mattingly}},
  \bibinfo{journal}{Phys. Rev.} \textbf{\bibinfo{volume}{D64}},
  \bibinfo{pages}{024028} (\bibinfo{year}{2001}).

\bibitem[{\citenamefont{Jacobson et~al.}(2003)\citenamefont{Jacobson, Liberati,
  and Mattingly}}]{Jacobson:2003ty}
\bibinfo{author}{\bibfnamefont{T.}~\bibnamefont{Jacobson}},
  \bibinfo{author}{\bibfnamefont{S.}~\bibnamefont{Liberati}}, \bibnamefont{and}
  \bibinfo{author}{\bibfnamefont{D.}~\bibnamefont{Mattingly}}
  (\bibinfo{year}{2003}), \eprint{gr-qc/0303001}.

\bibitem[{\citenamefont{Mattingly and Jacobson}(2001)}]{Mattingly:2001yd}
\bibinfo{author}{\bibfnamefont{D.}~\bibnamefont{Mattingly}} \bibnamefont{and}
  \bibinfo{author}{\bibfnamefont{T.}~\bibnamefont{Jacobson}}
  (\bibinfo{year}{2001}), \eprint{gr-qc/0112012}.

\bibitem[{\citenamefont{Kostelecky and Samuel}(1989)}]{Kostelecky:1989jw}
\bibinfo{author}{\bibfnamefont{V.~A.} \bibnamefont{Kostelecky}}
  \bibnamefont{and} \bibinfo{author}{\bibfnamefont{S.}~\bibnamefont{Samuel}},
  \bibinfo{journal}{Phys. Rev.} \textbf{\bibinfo{volume}{D40}},
  \bibinfo{pages}{1886} (\bibinfo{year}{1989}).

\bibitem[{\citenamefont{Kostelecky and Mewes}(2002)}]{Kostelecky:2002hh}
\bibinfo{author}{\bibfnamefont{V.~A.} \bibnamefont{Kostelecky}}
  \bibnamefont{and} \bibinfo{author}{\bibfnamefont{M.}~\bibnamefont{Mewes}},
  \bibinfo{journal}{Phys. Rev.} \textbf{\bibinfo{volume}{D66}},
  \bibinfo{pages}{056005} (\bibinfo{year}{2002}), \eprint{hep-ph/0205211}.

\bibitem[{\citenamefont{Lehnert}(2003)}]{Lehnert:2003ue}
\bibinfo{author}{\bibfnamefont{R.}~\bibnamefont{Lehnert}}
  (\bibinfo{year}{2003}), \eprint{gr-qc/0304013}.

\bibitem[{\citenamefont{Barbero}(1996)}]{Barbero:1996ud}
\bibinfo{author}{\bibfnamefont{J.~F.} \bibnamefont{Barbero}},
  \bibinfo{journal}{Phys. Rev.} \textbf{\bibinfo{volume}{D54}},
  \bibinfo{pages}{1492} (\bibinfo{year}{1996}), \eprint{gr-qc/9605066}.

\bibitem[{\citenamefont{Wald}()}]{Wald:1984rg}
\bibinfo{author}{\bibfnamefont{R.~M.} \bibnamefont{Wald}},
\bibinfo{title}{\textsl{General Relativity}},
  \bibinfo{note}{The University of Chicago Press, USA (1984)}.

\end{thebibliography}

\end{document}